\documentclass[12pt]{article}
\usepackage{pdproc,epsfig} 

\newcommand{\cob}{CO$^5$BOLD}
\newcommand{\cobe}{CO$^5$BOLD }
\newcommand{\rsun}{{\rm R_\odot}}

\begin{document}

\renewcommand{\thefootnote}{\alph{footnote}}
  
\title{SOLAR NEUTRINOS AND THE SUN}
 
\author{ALDO M. SERENELLI}

\address{Instituto de Ciencias del Espacio (CSIC-IEEC), \\
Facultad de Ciencias, Campus UAB, Bellaterra, 08193, Spain
\\
 {\rm E-mail: aldos@ice.csic.es}}

\abstract{We present updated standard solar models (SSMs) that incorporate the latest
results for nuclear fusion rates, recently published. We show helioseismic results for high and low 
metallicity compositions and also for an alternative set of solar abundance, derived from 3D model 
atmospheres, which give intermediate results. For the high and low metallicity models, we show that current
solar neutrino data can not differentiate between models and that a measurement of the CNO fluxes is 
necessary to achieve that goal. A few additional implications of a hypothetical measurement of CNO neutrinos,
both in terms of solar and stellar physics, are discussed.}
   
\normalsize\baselineskip=15pt

\section{Introduction}

We present and discuss the results of our most up-to-date solar models. Part of this work is devoted to 
the solar abundance problem, that is, the mismatch between results of solar models and helioseismic 
inferences on 
solar structure when most the recent and sophisticated (using a 3D solar model atmosphere) determination 
of the solar composition is used. Results for neutrino fluxes are also presented. Together with a new 
analysis of
solar neutrino data that incorporates latest results by the Borexino experiment, these results show
that solar neutrino data can be equally well reproduced by solar models with both the high and low 
metallicity compositions. 
Currently, solar neutrinos can not yield information towards the solution of the solar abundance problem. 
However, this situation could be modified if a direct determination of CNO neutrinos is achieved. If such a 
measurement were done today, it would lead to a direct measurement of the central C+N content of the Sun 
where uncertainties from solar models and neutrino properties represent only 10\%. Possible implications of 
such a measurement are discussed both in terms of solar and stellar physics. Finally, we speculate with the
possibility that the early evolution of the Sun and the interaction with its protoplanetary disk can be 
constrained by determining the central content of C+N. This might be of relevance in studies of the evolution 
of protoplanetary disks and planet formation theories.

\section{Solar Models}

\subsection{Astrophysical Factors}\label{sec:astro}

A decade after the critical evaluation of the pp chain and CN cycle rates published
in the Solar Fusion I paper\cite{sfi} a new revision, Solar Fusion II\cite{sfii}, has 
established a new set of recommended values and uncertainties for the pp-chains and 
CNO-bicycle cross sections. Results in Solar Fusion II account for the large effort of the
nuclear physics community, both experimental and theoretical, during the last 10 
years. Astrophysical factors have been regularly updated in 
SSM\cite{bs05,ssm09} during this period. We now adopt as our standard the recommended 
values in Solar Fusion II; the most relevant changes with respect to the previous 
values used in SSM calculations\cite{ssm09} are given in Table~\ref{tab:nuc}. Changes 
in the central values of key reactions are modest: +2\% for S$_{11}$, -2\% for 
S$_{33}$, -5\% for S$_{17}$, +6\% for S$_{1,14}$. The uncertainties are larger, in 
some cases by up to a factor of 2, than previous values. This has implications for the 
theoretical 
uncertainties in the neutrino fluxes, although uncertainties in the solar composition 
continue to dominate in most cases.

\begin{table}[ht]
\caption{New (SFII) and previous values of relevant astrophysical factors.\label
{tab:nuc}}
\footnotesize
\vskip 2pt
\centerline{\begin{tabular}{||l|c|c||}
\hline \hline 
Reaction & SFII & Previous \\
& (keV-b) & (keV-b) \\
\hline 
${\rm S}_{11}$ & $4.01\times 10^{-22} (1\pm 0.010)$ & $3.94\times 10^{-22} (1\pm 0.004)$\\
${\rm S}_{33}$ & $5.21\times 10^3 (1\pm 0.052) $ & $5.4\times 10^3 (1\pm 0.06)$\\
${\rm S}_{34}$ & $0.56 (1 \pm 0.054) $ & $0.567 (1 \pm 0.03)$ \\
${\rm S}_{17}$ & $2.08\times 10^{-2} (1 \pm 0.077)$& $2.14\times 10^{-2} (1 \pm 0.038)$\\
${\rm S}_{1,14}$ & $1.66 (1 \pm 0.072) $ & $1.57 (1 \pm 0.08)$ \\
\hline
R(pep)/R(pp) & $\uparrow$ 2.5\% & --- \\
\hline \hline
\end{tabular}
}
\end{table}

\subsection{Solar Composition(s)}

Most of the results presented here are based on two different sets of solar 
abundances. One is that of Grevesse \& Sauval (1998), hereafter GS98\cite{gs98}; the 
other, that from Asplund et al. (2009), hereafter AGSS09\cite{agss09}. For both sets 
we adopt the meteoritic scale for all refractory elements; silicon is the anchor 
point between the photospheric and meteoric scales. The adopted solar abundances are 
of fundamental importance because the surface metal-to-hydrogen abundances ratio is a
constraint solar models have to fulfill. Many properties of the resulting model 
depend on the adopted solar metallicity, or composition. In particular, the 
metal-to-hydrogen mass fractions ratio in the solar surface is $(Z/X)_\odot=0.0229$ 
according to GS98, but only 0.0178 according to AGSS09. The large change is mostly 
the result of a strong reduction (about 30-40\%) in the CNO and Ne abundances. 

The difference between GS98 and AGSS09 has a profound impact on solar models that has 
been widely discussed in the literature, particularly in relation to structural 
quantities in the models that can be compared with results from helioseismology. We 
briefly review these results in \S~3.1. Several reasons lie behind the 
reduction of the solar CNO abundances in \cite{agss09} but the 
dominating effects are: use of a 3D-hydrodynamics model atmosphere, better selection
of spectral lines (identification of blends), detailed treatment of radiative 
transport in the line-formation modeling including non-local thermodynamic 
equilibrium for some elements. 

Caffau and collaborators (hereafter \cob\cite{cobold}) have also 
embarked in a similar task, the determination of solar abundances using 3D state-of-
the-art model atmospheres. Interestingly, although the underlying structure of the 
model atmospheres 
by both Asplund's and the \cobe groups are very similar, derived abundances for the 
CNO elements are not; results are summarized in Table~\ref{tab:compo} 
(error bars are included for CNO elements). Whereas  
differences arise from a number reasons, the selection of spectral lines used by 
each group for the abundance determinations seems to be the dominant one. The 
treatment of blends and line broadening, for example, introduces non-negligible
differences between the results of different groups.

We note the \cobe results lay between the GS98 and AGSS09 values, in fact, they
are consistent within errors with both of them. Unfortunately, \cobe abundances have 
not been obtained at the moment for all elements and have to be complemented with 
other sources\cite{lodders}; abundances for these elements are given in parenthesis. 

It is not among the goals of the present work to make a critical analysis of the 
goodness of the different solar abundance determinations present in the literature, 
neither to discuss the detailed results of solar models for each set of abundances. 
We content ourselves by adopting GS98 and AGSS09 as two different standards which
probably represent two extreme cases and on which to base our discussion. In addition, we 
briefly present helioseismic results for a solar model using the \cobe abundances.

\begin{table}[ht]
\caption{Solar compositions used in this work}\label{tab:compo}
\footnotesize
\vskip 2pt
\centerline{
\begin{tabular}{||l|c|c|c||}
\hline \hline
Element & GS98 & AGSS09 & CO$^5$BOLD \\
\hline
C & $8.52\pm0.06 $& 8.43$\pm$0.05 & 8.50$\pm$0.06 \\
N & $7.92\pm0.06 $ & 7.83$\pm$0.05 & 7.86$\pm$0.12 \\
O & $8.83\pm0.06 $  & 8.69$\pm$0.05 & 8.76$\pm$0.07 \\
Ne & 8.08  & 7.93 & (8.05) \\
Mg & 7.58  & 7.53 & (7.54) \\
Si & 7.56  & 7.51 & (7.52) \\
Ar & 6.40  & 6.40 & (6.50) \\
Fe & 7.50  & 7.45 & 7.52 \\
\hline
Z/X & 0.0229  & 0.0178 & (0.0209) \\
\hline \hline
\end{tabular}
}
\end{table}

\section{Results}

\subsection{Helioseismology}

\begin{figure}[ht]
\centerline{\mbox{\epsfig{figure=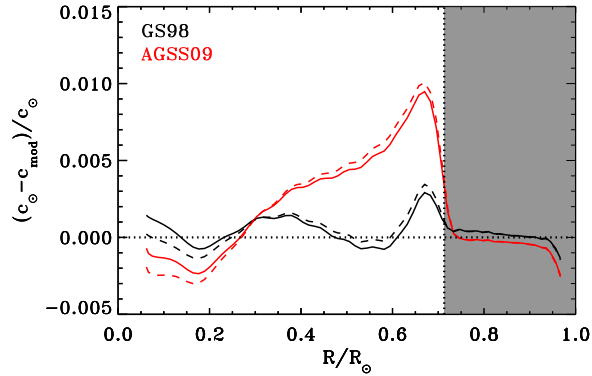,width=8.0cm}}}
\caption{Relative sound speed difference for SSMs with GS98 and AGSS09 compositions. 
Solid lines: new models incorporating SFII nuclear rates; dashed lines: previous 
generation of models. See text for more details.}
\label{fig:sound_sfii}
\end{figure}

The impact of solar abundances in 
the helioseismic properties of solar models has been widely discussed in the 
literature.\cite{montal04,bs05,guzik06,dela06,basu08} Solar models that use the 
AGSS09 composition are not in agreement with helioseismic inferences of the solar 
interior structure. Discrepancies manifest themselves in a variety of ways: mismatch 
in the determination of the depth of the convective envelope, low solar surface 
abundance of helium, differences in the sound and density profiles, too low mean 
molecular weight in the solar core. We suggest the interested reader to refer to the 
vast literature on the subject, a flavor of which is given in the references above, for 
details. Problems are present for standard solar models; they are even more acute 
for non-standard models that attempt to include poorly understood dynamic effects\cite{palacios:06}.
Most updated results for SSMs using our two reference compositions, GS98 and AGSS09, 
have been recently presented\cite{accre} and show small changes with respect to 
previous models. In 
Fig.~\ref{fig:sound_sfii} we show the relative sound speed difference for SSMs with
GS98 and AGSS09. For each composition, the dashed line shows results for previous 
models, i.e. models using the set of astrophysical factors given in the third column
of Table~\ref{tab:nuc}, whereas the solid line corresponds to models computed using
the Solar Fusion II recommended values. Changes are small but noticeable, 
particularly towards the center ($R<0.2\,\rsun$), and within 
the range of sensitivity of current helioseismic data\cite{basu2009}. 
The responsible for the changes is the slight increase in S$_{11}$ because it 
produces a decrease in the temperature of the central regions of the Sun. Since
for an ideal gas $c^2 \propto T / \mu$ (here $c$ is the sound speed, $T$ the 
temperature, and $\mu$ the mean molecular weight), and changes in $\mu$ are 
negligible, then the model sound speed is reduced accordingly. 

\begin{figure}[ht]
\centerline{\mbox{\epsfig{figure=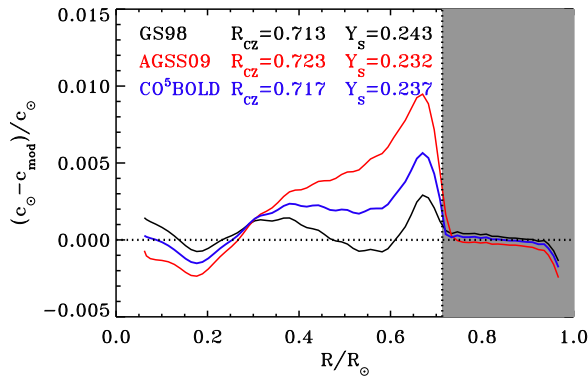,width=8.0cm}}}
\caption{Relative sound speed difference for SSMs using the different solar
abundances listed in Table~\ref{tab:compo}. Values for the depth of the convective envelope
in units of the solar radius and the mass fraction of the surface helium abundance are also shown.
For reference, the helioseismic values are $R{\rm_{CZ}/\rsun}=0.713\pm0.001$ and 
$Y_{\rm S}= 0.2485\pm0.0035$.}
\label{fig:sound}
\end{figure}

As can be expected from the abundances listed in Table~\ref{tab:compo}, when the
\cobe values are used, the resulting solar structure lays somewhere in between 
SSMs that use either GS98 or AGSS09. To illustrate this, the sound speed 
profile is given in Fig.~\ref{fig:sound}. The improvement in the sound speed 
profile model with  \cobe abundances with respect to the model with AGSS09 comes about 
because the location of the base of the convective envelope in the former is in better
agreement with helioseismology. Other helioseismic constraints, e.g. surface helium abundance, 
show a similar behavior, with \cobe abundances giving results almost equally distant from 
GS98 and AGSS09. Values for these quantities are included in Fig.~\ref{fig:sound}. 

\subsection{Solar Neutrinos}

The new SFII nuclear reaction rates do alter the
predicted solar neutrino fluxes, particularly the fluxes associated
with the ppIII chain and CN-cycle, mechanisms for solar hydrogen burning that
are relatively unimportant energetically. Changes in the nuclear rates in SFII with
respect to our previous set of preferred values have been described in 
\S~2.1. Model predictions for  the neutrino fluxes and associated uncertainties
are presented in
Table~\ref{tab:neutrinos} for both reference solar compositions, AGSS09 and GS98.  
The third column of the table quantifies the changes with respect to results using
older set of preferred nuclear rates \cite{ssm09}.
The most significant changes  are  a  5\% decrease  in  the
predicted $^8$B  flux primarily because of the increase in 
${\rm S_{11}}$ and  the increase in  the $^{13}$N flux  due to the  larger ${\rm S_{1,14}}$ 
and central abundance of C. The increase in C is a consequence of the lower SFII value
for $^{15}$N(p,$\gamma)^{16}$O, a reaction that competes with the CN I cycle
reaction $^{15}$N(p,$\alpha)^{12}$C and allows mass to flow out of the
CN I cycle into CN II.  

\begin{figure}[ht]
\centerline{\mbox{\epsfig{figure=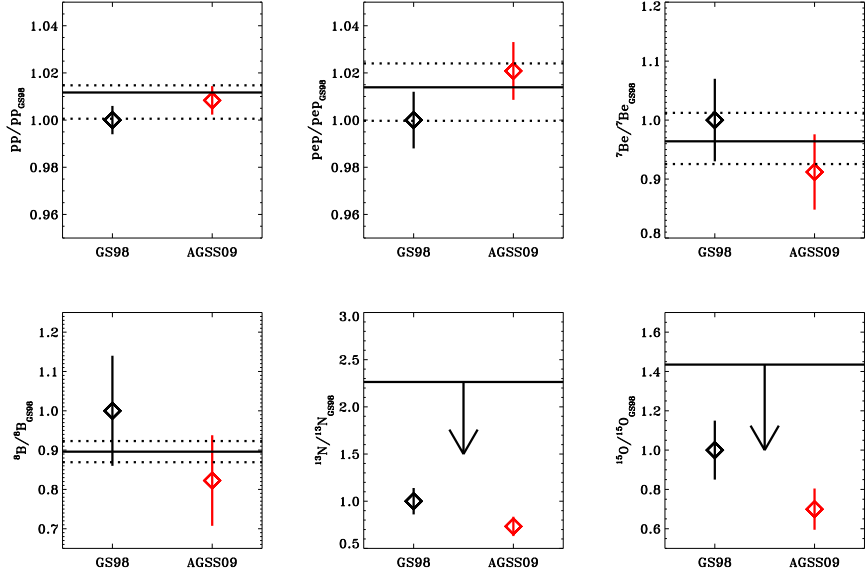,width=12.0cm}}}
\caption{Neutrino fluxes: comparison between experimental results and SSMs. Vertical
lines denote 1$\sigma$ model uncertainties. Solar values are represented by 
solid horizontal lines; dotted lines are 1$\sigma$ uncertainties. All fluxes 
normalized to GS98 SSM values.}
\label{fig:neutrinos}
\end{figure}

Table~\ref{tab:neutrinos}  also includes  the updated  solar  neutrino fluxes
inferred from all available neutrino  data\cite{accre}. The analysis includes the 
recent more precise  $^7$Be measurement\cite{borex:2011}, which is  the main change  
with respect to previous analysis \cite{roadmap,bps08,concha:2010}.  In Figure~\ref{fig:neutrinos} 
we show for the relevant neutrino fluxes a 
comparison between the two SSMs and solar fluxes, normalized to the GS98 SSM values. 
For the $^{13}$N and $^{15}$O fluxes only upper limits can be established with 
current solar neutrino data.  
Except for the pp flux, experimental values lay in between SSM predictions for the 
two reference solar compositions. 
For comparison of the SSM predictions with the fluxes inferred from neutrino data, we use 
the $\chi^2$ function defined in \cite{concha:2010}, with  updated errors   
and correlations. We find $\chi^2_{\rm GS98}=3.5$ and $\chi^2_{\rm AGSS09}=3.4$, 
leading in both cases to $P^{\rm agr}_{\rm GS98,AGSS09}=90\%$. The new fusion
cross sections from SFII and the new Borexino results  lead to  both models
predicting solar neutrino fluxes in  excellent  agreement  with  inferred
ones. From this analysis, we conclude that, currently, solar   neutrinos  can  not  
differentiate  between solar compositions. In both cases, excellent agreement with 
data is achieved. 

\begin{table}[ht]
\caption{Neutrino fluxes\label{tab:neutrinos}}
\footnotesize
\vskip 2pt
\centerline{
\begin{tabular}{||l|c|c|r|c||}
\hline \hline 
Flux & SFII-GS98 & SFII-AGSS09 & $\Delta$ & Solar\\
\hline
pp & $5.98 (1 \pm 0.006)$ & $6.03 (1 \pm 0.006)$ & +0.1\% &
$6.05 (1 ^{+0.003}_{-0.011})$ \\ 
pep & $1.44 (1 \pm 0.011)$ & $1.47 (1 \pm 0.012)$ & +2\% & 
$1.46 (1 ^{+0.010}_{-0.014})$ \\  
hep &  $8.04 (1 \pm 0.30)$ & $8.31 (1 \pm 0.30)$ & +1.6\% & $18 (1 ^{+0.4}_{-0.5})$\\ 
$^7$Be & $5.00 (1 \pm 0.07)$ & $4.56 (1 \pm 0.07)$ & -1.7\% & $4.82 (1 ^{+0.05}_
{-0.04})$\\ 
$^8$B  & $5.58 (1 \pm 0.14)$ & $4.59 (1 \pm 0.14)$ & -5\% & $5.00 (1 \pm 0.03)$ \\ 
$^{13}$N & $2.96 (1 \pm 0.14)$ & $2.17 (1 \pm 0.14)$ & +5\% & $\leq 6.7 $ \\ 
$^{15}$O & $2.23 (1 \pm 0.15)$ & $1.56 (1 \pm 0.15)$ & +5-6\% & $\leq 3.2 $\\ 
$^{17}$F & $5.52 (1 \pm 0.17)$ & $3.40 (1 \pm 0.16)$ & +2\% & $\leq 59 $\\
$\chi^2 / P^{\rm agr}$ & 3.5 / 90\% & 3.4 / 90\% & --- & --- \\
\hline
\hline
\end{tabular}
}
\end{table}

\section{CNO neutrinos: what can be learned?}

The Sun is almost entirely powered by proton-proton captures. However, nuclear 
energy generated by the CNO bi-cycle dominates for more massive ($1.2~{\rm M_\odot}$) 
or more evolved stars. Detection of solar CNO neutrinos is the only direct 
observational proof of CNO as a source of nuclear energy in stars. Current SSMs 
predict that the CNO bi-cycle contribute less than 1\% to the solar luminosity; 
CNO solar neutrinos can test this result.

If, on the other hand, we assume nuclear processes in stars are well understood, 
a measurement of CNO solar neutrinos can be used to reveal properties of the solar 
interior. In particular, $^{13}$N and $^{15}$O neutrinos have energies 
and fluxes which should be within the capabilities of current and forthcoming 
liquid scintillator detectors like Borexino, KamLAND, and SNO+. Recently, a new 
analysis technique for subtracting the $^{210}$Bi background from the Borexino signal
has been presented\cite{villante} that could allow detection of $^{13}$N and $^{15}$O
neutrinos in a Borexino-like detector with 1~yr of data. The prospects for 
precise measurements of CNO neutrinos are further improved with larger detectors like 
SNO+. 

A direct goal of measuring CNO neutrinos is that they can be used to determine 
abundance of carbon plus nitrogen in the solar core. As discussed above, current 
solar neutrino data not only does not discriminate between high and low-metallicity 
solar 
compositions but can not even favor one of them. Let us assume an optimistic scenario 
in which both the $\Phi({\rm ^{13}N})$ and $\Phi({\rm ^{15}O})$ fluxes are measured 
to 10\% precision. Moreover, let the central values perfectly align with results 
from one 
of the SSMs discussed above. If the GS98 SSMs values are used, i.e. 
$\Phi_{\rm exp}({\rm ^{13}N})=2.96\times 10^8{\rm cm^{-2} s^{-1}}$ and 
$\Phi_{\rm exp}({\rm ^{15}O})=2.17\times 10^8{\rm cm^{-2} s^{-1}}$, then we would
get $\chi^2_{\rm GS98}= 3.7 (88\%)$ and $\chi^2_{\rm AGSS09}= 13.9 (8\%)$; i.e.
measuring $\Phi({\rm ^{13}N})$ and $\Phi({\rm ^{15}O})$ at this level of precision
would clearly favor solar models with a particular C+N content over the other. 
Clearly, these assumptions make up for a very favorable case, but give a flavor of 
what can potentially be achieved.

A more targeted approach can be used to determine the C+N abundance in the solar 
core directly. One possible way has been laid out where the $^8$B solar neutrino flux 
is used as a thermometer of the solar core\cite{cno}. Because the temperature 
dependence of the nuclear reactions that produce CNO and $^8$B neutrinos is very 
similar, many sources of uncertainty in the solar models affect them in the same way 
and can be almost completely cancelled out; these are the {\em environmental 
factors}. Using power-law expansions of solar neutrino fluxes\cite{bahcall89}, 
$\Phi({\rm ^{13}N}+{\rm ^{15}O})$ can be expressed as a function of 
$\Phi({\rm ^8B})$ as\cite{cno}:
\begin{eqnarray}\label{eq:cno}
\frac{\Phi({\rm ^{13}N}+{\rm ^{15}O})}{\Phi^{\rm SSM}({\rm ^{13}N}+{\rm ^{15}O})} = 
\frac{X(C+N)}{X^{\rm SSM}(C+N)}\left[\frac{\Phi({\rm ^8B)}}{\Phi^{\rm SSM}({\rm 
^8B})}\right]^{0.828} \times \ \ \ \ \ \ \ \  \ \ \ \ \ \ \ \ \ \ \ \ \ \ \ \ \\
\left[{\rm 1\pm 0.03(exp) \pm 0.026(env) \pm
0.035(LMA) \pm 0.10(nucl)}\right]. \nonumber
\end{eqnarray}
Uncertainties in this equation are: 3\% from the experimental $^8$B measurements, 
2.6\% from remainder of environmental uncertainties, 3.5\% from neutrino parameters, 
and 10\% from nuclear cross sections. These uncertainty sources are experimental and 
under control, and can be eventually improved (except for the negligible 
environmental component). The SSM only acts as a scaling factor (and determining the 
precise value of the power-law exponent, but this changes little for different solar 
models). A hypothetical measurement of $\Phi({\rm ^{13}N}+{\rm ^{15}O})$ can directly 
yield a determination of the central C+N abundance. The uncertainty from all other 
sources besides the hypothetical measurement amounts, today, to $\sim 10\%$ and are dominated 
by the nuclear
physics part. For reference, the difference in the C+N abundance brought about by the change 
from the GS98 to the AGSS09 solar abundances is of the order 35\%.

There is not direct information about the composition of the solar (or 
any other star) core. Helioseismic information about the solar core is mostly 
sensitive to the mean molecular weight and temperature; information about metallicity is of
indirect nature, and usually degenerate with other quantities, such as the radiative opacity
of stellar matter. CNO neutrinos would give direct information 
on the abundance of metals, in particular C+N, in the solar central regions. This 
would be in itself an outstanding achievement. Together with other constraints on the 
solar composition, this could potentially be used to constrain, among other 
processes, the efficiency of settling of heavy elements in the Sun. 

Recently, it has been suggested that there are systematic differences between
the solar surface abundances and those of other stars, otherwise very similar to the 
Sun, depending on the presence and characteristics of the planetary system in each 
star\cite{twins1}. Tentatively, the cause for these differences has been ascribed to 
an interplay between refractory elements been preferentially locked in planetesimals
and protoplanetary disk material been accreted to the Sun. If true, this idea would
suggest the surface composition of the Sun does not reflect the original composition 
of the protosolar nebula but rather it is the outcome of a mixture between 
composition of the  primordial nebula and of chemically processed material from the 
protoplanetary disk. The solar core, however, keeps memory of the primordial 
composition. 

\begin{figure}[ht]
\centerline{\mbox{\epsfig{figure=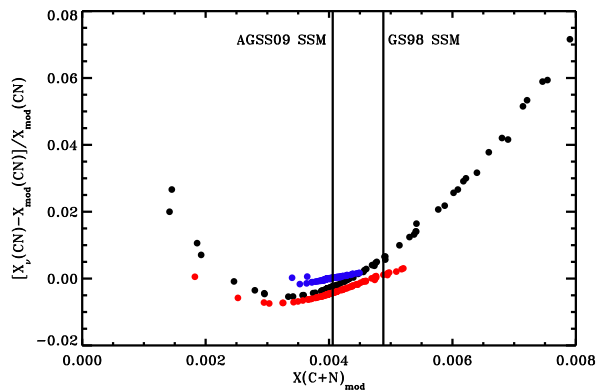,width=8.0cm}}}
\caption{Neutrino fluxes: comparison between experimental results and SSMs. Vertical
lines denote 1$\sigma$ model uncertainties. Solar values are represented by 
solid horizontal lines; dotted lines are 1$\sigma$ uncertainties. All fluxes 
normalized to GS98 SSM values.}
\label{fig:cn}
\end{figure}

Because CNO neutrinos offer a unique way of determining solar core metal content
(at least of C and N), the interesting possibility arises they can 
be used to constraint the evolution and interaction between the early Sun and 
its protoplanetary disk. A first step towards assessing the possibility the Sun has 
suffered such an accretion phase has been 
recently discussed\cite{accre}. In that work different accretion histories (varying 
accreted masses and compositions) have been considered and the resulting structure of 
solar models, both in helioseismic and solar neutrino aspects, analyzed. For reasons 
of space, we can not summarize those results here. However, in terms of our present discussion, 
possibly the most interesting result is that applying Eq.~\ref{eq:cno} (developed for 
SSMs) to the non-SSMs that include accretion, the central C+N abundance of the models
can be recovered with very good precision. Using the $^8$B, $^{13}$N, and $^{15}$O 
fluxes for each of the models with accretion, we can use Eq.~\ref{eq:cno} to estimate the central 
content of C+N in the model, ${\rm X_\nu(CN)}$. The relative difference between this estimation
and the actual C+N central abundance of the model, ${\rm X_{mod} (CN)}$, is shown as a function
of ${\rm X_{mod} (CN)}$ in Fig.~\ref{fig:cn}. For illustration, results for the GS98 and AGSS09 SSMs 
are shown with black vertical lines. Neutrino fluxes {\em alone} allow the determination 
of the central C+N abundance with 6\% accuracy for all models considered which, as can be seen, have
quite large variations in the metallicity. This is better than the current intrinsic uncertainty
in the relation, 10\%, dominated by uncertainties in astrophysical factors ($S_{\rm 1,14}$ and 
$S_{\rm 17}$). Results, therefore, are encouraging: scaling relations between neutrino fluxes
and solar composition derived from SSMs seem to be applicable in quite general cases, even for 
cases where the central metallicity changes by more than a factor of 2, above of what can 
be reasonably expected given helioseismic and current solar neutrino constraints\cite{accre}. 

\section{Final remarks}

Latest changes in the input physics of SSMs have a small effect on helioseismic quantities 
and do not have an impact on the solar abundance problem. A comparison of the new analysis of solar 
neutrino data, including the latest results by Borexino, against model predictions show that
current neutrino data can be equally described by SSMs with high and low metallicity compositions. Both
experimental $^8$B and $^7$Be fluxes lay right in between both model predictions for both GS98 and AGSS09
compositions. On the other hand, a precise measurement of the $^{13}$N and $^{15}$O fluxes allows to extract 
the central C+N abundance with a model uncertainty of about 10\%. To the extent we have been able to test this 
result, it seems valid regardless of SSMs being an accurate representation of the solar structure. 
A measurement of the CNO fluxes, and as a corollary of the central C+N abundance, has a wide range of
implications not only for solar but also for stellar physics; here we have briefly discussed a few of them:
direct experimental determination of CNO-cycles as source of nuclear energy in stars, solar abundance problem, 
mixing processes in the Sun and, more tentatively, early phases of solar evolution and interaction with the
protoplanetary disk.

\section{Acknowledgements}
I am grateful to the organizers of NEUTEL~11 for the invitation to participate in 
such a stimulating workshop.
This work is  partially supported  by the  European  Union International
Reintegration Grant  PIRG-GA-2009-247732, the MICINN  grant AYA08-1839/ESP, by
the  ESF EUROCORES Program  EuroGENESIS (MICINN  grant EUI2009-04170),  by SGR
grants  of   the  Generalitat  de   Catalunya  and  by  the   EU-FEDER  funds.

\end{document}